\documentclass[twocolumn,prl,superscriptaddress,floatfix,showpacs]{revtex4-1}
\usepackage[utf8]{inputenc}
\usepackage{amsmath}
\usepackage{amssymb}
\usepackage{graphicx}

\makeatletter
\usepackage{graphics}
\usepackage{epsfig}
\usepackage{color}

\newcommand{\be}{\begin{equation}} \newcommand{\ee}{\end{equation}}
\newcommand{\bea}{\begin{eqnarray}} \newcommand{\eea}{\end{eqnarray}}

\makeatother

\begin{document}

\title{
Neuronal cable equations derived from the hydrodynamic motion of charged particles}

\author{Davide Forcella} 
\affiliation{Paris-Saclay University, CNRS, Gif sur Yvette, France}
\affiliation{The European Institute of Theoretical Neuroscience (EITN), Paris, France}
\author{Alberto Romagnoni} 
\affiliation{Paris-Saclay University, CNRS, Gif sur Yvette, France}
\affiliation{The European Institute of Theoretical Neuroscience (EITN), Paris, France}
\author{Alain Destexhe} 
\affiliation{Paris-Saclay University, CNRS, Gif sur Yvette, France}
\affiliation{The European Institute of Theoretical Neuroscience (EITN), Paris, France}

\date{\today}

\begin{abstract}

Neuronal cable theory is usually derived from an electric analogue of the membrane, which contrasts with the slow movement of ions in aqueous media.  We show here that it is possible to derive neuronal cable equations from a different perspective, based on the laws of hydrodynamic motion of charged particles (Navier-Stokes equations).  This results in similar cable equations, but with additional contributions arising from nonlinear interactions inherent to fluid dynamics, and which may shape the integrative properties of the neurons.

\end{abstract}

\maketitle

The dynamics of the membrane potential (V$_m$) of neurons in their
complex dendritic arborization can be described by cable equations,
introduced by Rall almost 60 years ago~\cite{Rall62} and which are
still used today.  These equations are based on a set
of partial-differential equations (PDEs) that describe the V$_m$
dynamics in space and time.  The derivation of cable equations is
usually based on an electric-circuit analogue of the membrane,
which is widely used to model neurons~\cite{Hille,HH}.  This
formalism has been very successful to model a wide range of neural phenomena
involving dendrites~\cite{Rall95}.

Besides this success, treating the neuron as an
electric circuit has some drawbacks. For example, the classic cable theory
accounts for charge accumulation in the membrane capacitance, but forbids
charge accumulation inside the cable, and thus forbids the formation of
electric monopoles in neurons~\cite{commentary2012}.  Ideally, one should
derive equivalent cable equations by allowing dendritic charge accumulation, which was
attempted through the ``generalized cable'' model~\cite{BedDes2013}.
In that approach, the current is extended to include the
displacement current, thus allowing charge accumulation inside the dendrite.  However, the
generalized cable is still based on an electric circuit analogue. 
As a consequence, the {electromagnetic signal} is considered as instantaneous, 
which appears non realistic for neurons.  Indeed, in neuronal cables,
the charges are ions moving in 
aqueous media (cytoplasm or extracellular fluid), and have a mobility orders of magnitude lower
than electrons in metal. For instance, the mobility of electrons in copper is of $4.45\times10^{-3}~m^2/sV$ for a temperature of $298.15~ ^oK$ \cite{PhilipBolton2002}, which
is about $10^5$ times larger than the mobility of Na$^+$ in sea water, which is of $5.19\times 10^{-8}~m^2/sV$, and is similar for other ions such as
K$^+$, Ca$^{2+}$ and Cl$^{-}$ \cite{Hille}.  Despite such huge differences,
there is presently no formalism to correctly describe this slow charge
movement.

In the present letter, we show that it is possible to derive cable
equations from another perspective, which does not require to make the
assumptions of instantaneity of electric circuits.  We treat ions
moving in aqueous media as fluids, and use the equations of
hydrodynamic motion to describe this flow.  We show that, considering
this fluid as linear, leads to the classic passive cable equations.
Including the nonlinearity inherent to fluids leads to novel forms of
nonlinearity, that could be tested experimentally.



The classic derivation of the cable equation starts from Ohm's law,
according to which the axial current $i_i$ in a cylindric cable can be
written as:
\begin{equation}
    i_i = -\frac{1}{r_i}\frac{\partial V_m}{\partial x}~,
\end{equation}
where $V_m$ is the membrane potential, $r_i$ is the resistivity in the
axial direction, and $x$ is the distance along the axis of the cylindric
cable.  The current balance implies that the variation of the current
along $x$ is equal to the membrane current $i_m$:
\begin{equation}
    i_m = -\frac{\partial i_i}{\partial x} ~ .
\end{equation}
According to the RC-circuit analogue of the membrane, the membrane
current is given by:
\begin{equation}
    i_m=c_m \frac{\partial V_m}{\partial t} +\frac{V_m}{r_m} ~ ,
\end{equation}
where $c_m$ is the specific membrane capacitance, and $r_m$ the
resistivity of the membrane.  Combining the above equations, we obtain:
\begin{equation}
\label{eqA}
\frac{1}{r_i} \frac{\partial^2 V_m}{\partial x^2} 
  - c_m \frac{\partial V_m}{\partial t} - \frac{V_m}{r_m}=0 ~ ,
\end{equation}
which is known as the cable equation.  Note that, in this form, the
cable is of constant radius, and $c_m$, $r_m$, and $r_i$ are constant.

Using the more compact notation $\partial_t$ and $\partial_x$ for the partial derivatives, 
we can rewrite the cable equation in
the form:
\begin{equation}
\label{eqB}
\Big( \tau_{CT} \partial_t - \lambda_{CT}^2 \partial_x^2 + \mathbb{I} \Big)~ V_m=0 ,
\end{equation}
where the length and time constants are:
\begin{eqnarray}
&& \lambda_{CT} = \sqrt{\frac{r_m}{r_i}}  \\
&& \tau_{CT} = r_m c_m.
\end{eqnarray}
Since the operator acting on $V_m$ in Eq. \ref{eqB} commutes with spatial and temporal derivatives, we can rewrite the equation for the electric current vector $\vec{i}$, 
\begin{equation}
\label{eqB2}
\Big( \tau_{CT} \partial_t - \lambda_{CT}^2 \partial_x^2 + \mathbb{I} \Big)~ \vec{i}=0 ,
\end{equation}
where $i_x=i_i$ is the axial current, and $i_y=i_z= \kappa~ i_m$, with $i_m$ the membrane current and $\kappa$ a suitable constant with the dimension of a length. Notice that this would no longer be true if $\kappa$ does depend on $t$ and $x$.


To describe neuronal cables from a different perspective, we consider
the flow of ions into the aqueous medium of dendrites and axons
as the flow of a charged fluid subject to boundary conditions at the border (membrane) 
\cite{Forcella}.  
A large class of fluids can be described by very well-known
Navier-Stokes (NS) equations \cite{Landau}:
\begin{eqnarray}
       \rho \Big( \partial_t \vec{v} + ( \vec{v}\cdot \vec{\nabla} ) \vec{v}  \Big) & = & 
       - \vec{\nabla} P+ \eta \nabla^2  \vec{v} 
       + \Big( \frac{1}{3} \eta  +\zeta \Big) \vec{\nabla} ( \vec{\nabla} \cdot \vec{v} )  
       \nonumber \\
       & & +  \vec{f} - \rho \frac{ \vec{v}}{\tau_{NS}}
    \label{eqC}
\end{eqnarray}
where $\vec{v}$ is the velocity vector field of the fluid, $\rho$ is
the density of the fluid (for the fluid at equilibrium: constant and
homogeneous), $P$ is the external pressure, $\eta$ the shear viscosity
of the fluid, while $\zeta$ is its bulk viscosity, and $\vec{f}$ the
set of vector forces external to the fluid (e.g. due to external
electrical field, gravity, etc.). The term $\frac{ \rho \vec{v}}{\tau_{NS}}$ is
a correction to the NS equations due to interaction of the fluid with
the external medium: it is an effective friction term that causes the
slowdown the flow of the fluid. It is a first order correction to
NS equations, where $\tau_{NS}$ is the mean free time of
interaction between fluid's components and the external medium.


For a charged fluid in a neuronal cable, we
consider the following conditions:
\begin{itemize}
    \item in the absence of external forces, $\vec{f} = 0$;
    \item with a homogeneous pressure, $\vec{\nabla} P = 0$;
    \item the fluid is incompressible, $\vec{\nabla} \cdot \vec{v}=0$;
    \item the fluid is linear, $ ( \vec{v}\cdot \vec{\nabla} )~\vec{v} \sim 0 $;
\end{itemize}
Under these hypothesis, the NS Eq. (\ref{eqC}) then reduces to:
\begin{equation}
\label{eqD}
\Big( \rho \partial_t - \eta \nabla^2  + \frac{\rho}{\tau_{NS}}\Big) ~\vec{v}=0.
\end{equation}
The current density is naturally defined as $\vec{j} =\rho_q \vec{v}$, where the average charge density is $\rho_q = q \rho$, and $q$ is the average charge carried by an element of fluid. Note that $q$ should be considered as a ``net'' charge, because both positive and negative ions contribute to the charged fluid (this is similar to consider the total membrane current although it is carried by different ions).
In the case of constant and homogeneous charge density ($\partial_t
\rho_q = \vec{\nabla} \rho_q =0$), or if the charge density satisfies the
diffusion equation $\partial_t \rho_q - \frac{ \eta}{\rho} \nabla^2
\rho_q =0$, Eq. (\ref{eqD}) can be written as:
\begin{equation}
\label{eqE}
\Big( \tau_{NS} \partial_t - \lambda^2_{NS} \nabla^2  + \mathbb{I} \Big) ~\vec{j}=0,
\end{equation}
where $\lambda^2_{NS} = \frac{\tau_{NS} \eta}{\rho}$. This equation has exactly the same mathematical form as the cable Eq. (\ref{eqB}). 

Notice that in particular this duality implies the correspondence $\lambda^2_{CT} = \lambda^2_{NS}$ and $ \tau_{CT} = \tau_{NS}$, which gives the following relations:
\begin{eqnarray}
    && \eta= \frac{\rho}{r_i c_m}, \\
    && \tau_{NS} = r_m c_m
\end{eqnarray}
Note that this equality holds only in the case of a uniformly charged fluid where the charge density $q \rho$ is independent of position and time.  In reality, most of the ions are located close to the neuronal membrane~\cite{Hille}, so the charge density is non-uniform.  Nevertheless, in a one-dimensional approximation of the cable (neglecting radial variations of density), and in cables of constant diameter (neglecting longitudinal changes in density), this relation should be useful to relate hydrodynamic and neuronal variables.  Also note that if transient input currents are located along the dendritic cable (such as synapses), the charge density will change locally, possibly forming transient monopoles before re-equilibration.


Thus, we see that a form equivalent to cable equations 
can be derived from the linearized version of Navier-Stokes equations.  
However, Navier-Stokes equations are intrinsically non-linear, and can be approximated by linear equations only in specific circumstances. If we now consider the full Navier-Stokes equations (Eqs.~\ref{eqC}) without linear approximation, thus taking into account the nonlinear term $\rho(\vec{v}\cdot \vec{\nabla} ) \vec{v}$, provides a natural 
non-linear correction to cable equations (\ref{eqB}):
\begin{equation}
\label{eqF}
\Big( \partial_t - \frac{ \eta}{\rho} \nabla^2  + \frac{ 1}{\rho \tau}\Big) \vec{j} + \frac{1}{q \rho} (\vec{j}\cdot \vec{\nabla} ) \vec{j} =0
\end{equation}
In the linear response regime: $\vec{j}= \frac{1}{\vec{r}}\vec{E}$ with $\vec{r}$ the vector of resistivity of the medium: $r_x=r_i$, $r_y=r_z=r_m$ 
In the simple case in which $\vec{E}= - \vec{\nabla} V_m $, equation (\ref{eqF}) becomes: 
\begin{equation}
\label{eqG}
\Big( \partial_t - \frac{ \eta}{\rho} \nabla^2  + \frac{ 1}{\rho \tau}\Big) V_m - \frac{1}{ q \rho } (  \frac{\vec{\nabla}}{\vec{r} }  V_m \cdot \vec{\nabla} ) V_m =0
\end{equation}
Reintroducing the classical coefficients used in cable equation (\ref{eqA}), and considering the particular case in which $V_m$ depends only on $x$, leads to: 
\begin{equation}
\label{eqH}
\Big( - c_m\partial_t V_m  + \frac{1}{r_i} \partial_x^2- \frac{1}{r_m}\Big) V_m + \frac{1}{q r_i} (\partial_x V_m)^2 =0
\end{equation}

Thus, a new nonlinear term $\frac{1}{q r_i} (\partial_x V_m)^2$ appears in cable equations, due to the nonlinear nature of fluid dynamics as described by Navier-Stokes equations. 


To conclude, we have shown here that the well-known cable equations of neurons~\cite{Rall62} can be derived from fluid dynamics considerations.  The cable equations are classically derived from the RC-circuit analogue of the membrane.  We show here that very similar cable equations can be derived without associating the cable membrane as a series of RC circuits, but rather from the hydrodynamic motion of charged particles.  We discuss below the possible consequences of this work, and openings for future studies.

A first consequence is that there is no need to necessarily associate an RC circuit to the cable interactions in neuronal structures.  The RC circuit describes well the ionic and capacitive currents in neuronal membranes.  However, associating a RC circuit analogue to the current flow (axial current) inside the dendrites makes the assumption that {the electromagnetic signal propagates} infinitely fast.  Such an assumption may be good to describe electronic circuits, but biological media are much slower, principally because the charges are ions moving in a fluid (cytoplasm or extracellular space), which have a mobility 5 orders of magnitude less than electrons in a metal~\cite{PhilipBolton2002}.  Thus, as noticed before~\cite{commentary2012}, the RC-circuit analogue forbids phenomena such as charge accumulation inside the dendrite, {although there is experimental evidence for such charge accumulation and electric monopoles in neural tissue}~\cite{Riera}.  With the hydrodynamic analogue we propose here, such charge accumulation would a priori be possible, because fluid dynamics is fully compatible with slow charge movement.  This aspect should be examined in more detail in future studies.

A second consequence of this hydrodynamic analogue is that going away from the linear approximation leads to a new term in cable equations.  This constitutes a strong prediction of this formalism.  
Notice that in general, two different types of corrections to linear approximation of NS equations for a fluid could be taken into account: corrections proportional to powers of  $\vec{v}$ and corrections proportional to powers of  derivatives of $\vec{v}$. The first type are due to interactions between the fluid and the external medium. Those are terms of friction which break translational invariance. The linear term $\sim \frac{\vec{v}}{\tau_{NS}}$ breaks the invariance in an isotropic way, and $\tau_{NS}$ is the thermalization time of this type of interactions. On the other hand, the second type of correction, which we are considering here, is related to internal fluid-fluid interactions and are the first non-linear corrections to the classical linear approximation of NS. 

Future studies should examine possible consequences of this additional nonlinear term, as well as the terms which arise from the relaxation of the constant and homogeneous charge density hypothesis, on the integrative properties of neurons.  Further, one should also examine if some of these consequences could be measured experimentally, resulting in a test of the predictions of the present formalism.  

The hydrodynamic cable formalism will allow us to investigate the functional consequences of the slow movement of charges in neurons, which was previously forbidden by describing neurons as (infinitely fast) electric circuits.  More generally, it provides a first step towards a description of the slow movement of charges inherent to biological media.

\begin{acknowledgments}

  We thank Claude Bedard for useful discussions.  Research funded by
  the CNRS, the European Community (H2020-720270, H2020-785907), the
  ANR (PARADOX) and the ICODE excellence network.

\end{acknowledgments}


\end{document}